\documentclass[journal=nalefd,manuscript=letter,layout=twocolumn]{achemso}
\usepackage{chemformula} % Formula subscripts using \ch{}
\usepackage[T1]{fontenc} % Use modern font encodings
\usepackage[utf8]{inputenc}
\usepackage{multicol}
\pdfoutput=1
\graphicspath{Figures}
\author{Samantha Sbarra}
\affiliation{Matériaux et Phénomènes Quantiques, Université de Paris, CNRS, UMR 7162, 10 rue Alice Domon et Léonie Duquet, Paris 75013, France}

\author{Louis Waquier}
\affiliation{Matériaux et Phénomènes Quantiques, Université de Paris, CNRS, UMR 7162, 10 rue Alice Domon et Léonie Duquet, Paris 75013, France}

\author{Stephan Suffit}
\affiliation{Matériaux et Phénomènes Quantiques, Université de Paris, CNRS, UMR 7162, 10 rue Alice Domon et Léonie Duquet, Paris 75013, France}

\author{Aristide Lemaître}
\affiliation{Centre de Nanosciences et de Nanotechnologies, CNRS, UMR 9001, Université Paris-Saclay, Palaiseau 91120, France}

\author{Ivan Favero}
\affiliation{Matériaux et Phénomènes Quantiques, Université de Paris, CNRS, UMR 7162, 10 rue Alice Domon et Léonie Duquet, Paris 75013, France}
\email{ivan.favero@u-paris.fr}

\title{Multimode optomechanical weighing of a single nanoparticle}
\begin{document}
\begin{abstract}
We demonstrate multimode optomechanical sensing of individual nanoparticles with radius of a hundred of nanometers. A semiconductor optomechanical disk resonator is optically driven and detected under ambient conditions, as nebulized nanoparticles land on it. Multiple mechanical and optical resonant signals of the disk are tracked simultaneously, providing access to several physical informations about the landing analyte in real-time. Thanks to a fast camera registering the time and position of landing, these signals can be employed to weigh each nanoparticle with precision. Sources of error and deviation are discussed and modeled, indicating a path to evaluate the elasticity of the nanoparticles on top of their mere mass. The device is optimized for future investigation of biological particles in the high megadalton range, such as large viruses.
\end{abstract}

\section{Introduction}
The potential of nanomechanical devices for mass sensing has been well illustrated in the literature \cite{Chaste2012, Hanay2012}. Their small size enables great performances for masses between the megadalton and several gigadaltons, which remain poorly accessible to other conventional mass spectrometry techniques. This mass range is of utmost biological interest, hosting entities such as viruses and bacteria, and first promising steps were indeed taken to weigh capsides with nanomechanical resonating strings, whose motion was read-out by electrical means \cite{HentzCapsid2018}. Over the years, nanomechanical mass sensors have taken on a variety of geometries and adopted several actuation and read-out techniques. Amongst them, all-optical techniques bring about a wideband and low-noise actuation capacity \cite{Sauer2017,Allain2020,Guha2020}, together with an outstanding sensitivity to mechanical motion, which are central advantages when working with small/high-frequency resonators. Such optical techniques naturally benefit from optomechanical concepts \cite{Favero2009,Aspelmeyer2014} and miniature optomechanical resonators were recently investigated for the mechanical sensing of objects of nanoscale mass \cite{Liu2013,Yu2016,Venkatasubramanian2016,Maksymowych2019,Gil-Santos2020,Sansa2020a}, with an improving level of control. This culminated in ref \citenum{Sansa2020a}, where single particle nanomechanical mass spectrometry in the megadalton range was demonstrated using an optomechanical nano-ram device with capture area of $4.5$ $\mu$m$^{2}$. A single mechanical mode of the nano-ram was operated in vacuum, and transduced through coupling with the optical mode of a separated ring resonator.

Another route may consist in using semiconductor disk optomechanical resonators \cite{Ding2010} to weigh individual nano-objects. In these disks, mechanical and optical modes are in contrast co-localized, which opens the possibility for dual mechanical and optical sensing of the deposited analyte. The co-localization also induces an intense coupling between light and motion \cite{Eichenfield2009, Baker2014b}, which enables efficient optical actuation and read-out of multiple mechanical modes at the same time \cite{Sbarra2021}, providing a path towards multimode mechanical sensing. Finally, the in-plane vibrations of disk resonators are little affected by viscous damping, ensuring high-level performances both in air and liquid environment \cite{Gil-Santos2015,Fong2015,Hermouet2019}. We take advantage of all these assets here and report on the optomechanical weighing of individual nanoparticles by a gallium arsenide (GaAs) disk resonator operated optically in ambient air environment. The disk combines an experimental detection limit of 40~MDa with a capture area of $380$ $\mu$m$^{2}$, two decades above recent realizations \cite{Venkatasubramanian2016,Maksymowych2019,Sansa2020a}. We demonstrate dual mechanical and optical sensing, as well as multimode mechanical sensing, obtaining multiple concomitant informations on the deposited particles in real-time. The optomechanical device is optimized to weigh masses up to tens of GDa, and we are able to measure and quantitatively analyse the deposition of individual nanoparticles that mimick intermediate viruses, be it by their size and mass (150 nm diameter nanoparticles), or by their elastic modulus (soft latex nanoparticles).

\section{Experimental section}
	\begin{figure*}[h!]
	\includegraphics[width=0.8\textwidth]{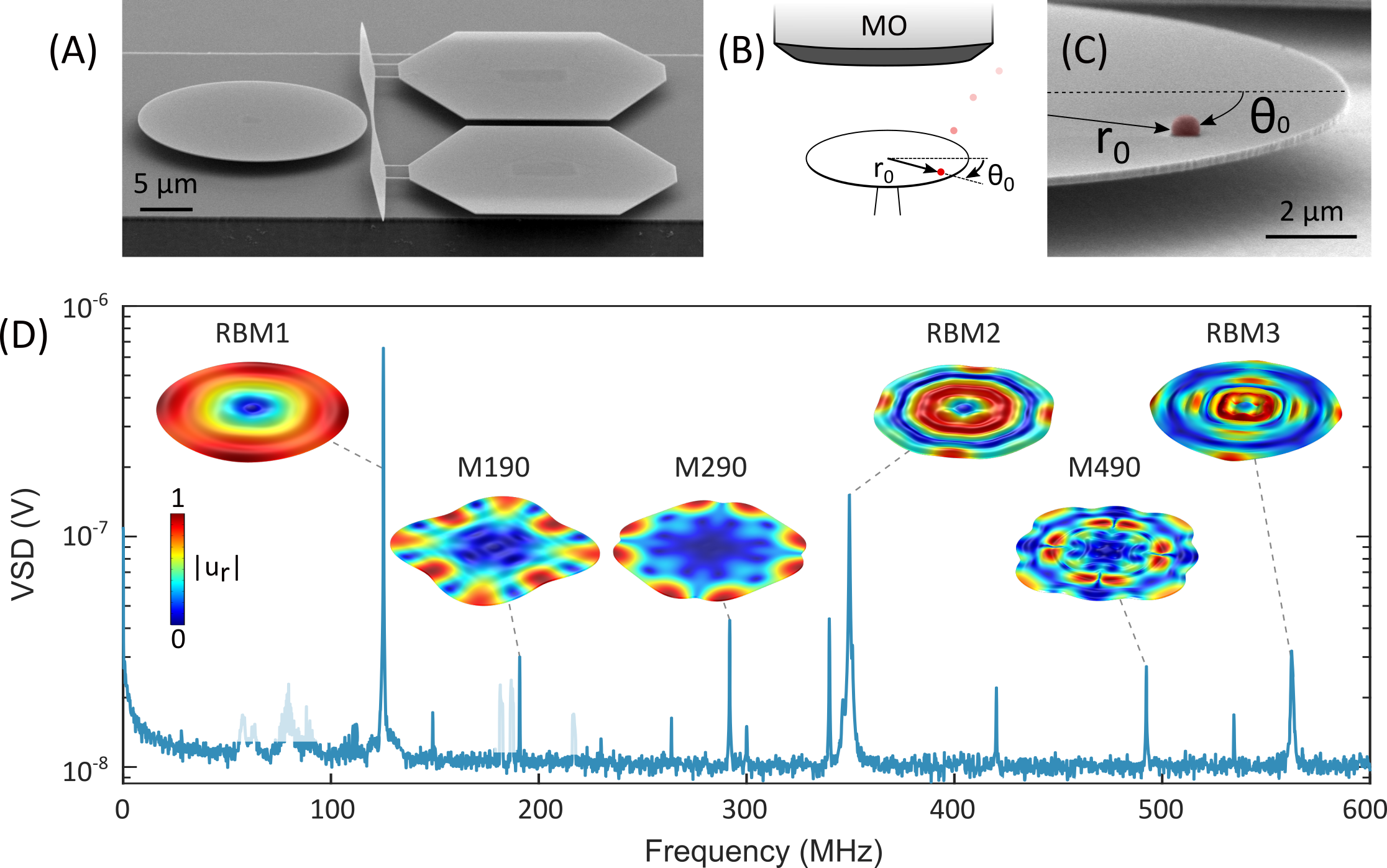}
	\caption{(A) Electron micrograph of the optomechanical disk resonator employed for nanoparticle sensing (left), together with its coupling nanowaveguide (middle) and its two anchoring pads (right). (B) A microscope objective (MO) installed in the set-up allows estimating the nanoparticle (red dot) landing position. (C) A finer measurement of the radial (r$_0$) and azimuthal ($\theta_0$) nanoparticle position is obtained through SEM imaging. $\theta_0$ is the azimuthal angle with respect to the <100> crystalline axis. (D) The Brownian motion of several mechanical modes appears in the RF spectrum of the disk optical output. The related mechanical modes profiles are shown in inset. Matted peaks in the spectrum are related to electronic instrumental noise.}
	\label{fig:Figure1}
	\end{figure*}

Our optomechanical sensor, shown in Figure~\ref{fig:Figure1}A, consists of a GaAs disk of 11~$\mu$m radius and 200~nm thickness sitting on a Aluminium Gallium Arsenide (AlGaAs) pedestal. A suspended optical nanowaveguide hold by two hexagonal anchoring pads is positioned in the resonator vicinity to allow evanescent laser light injection and collection into and from the disk \cite{BakerWaveguide2011}. By sinusoidally modulating the telecom laser at a radio-frequency close to mechanical resonances, the disk mechanical motion is actuated and information about the mechanical amplitude and phase is imprinted on the output optical signal and analyzed by demodulation \cite{Sbarra2021}. This all-optical modulation/demodulation scheme, combined with a proper understanding of the demodulated signal \cite{Sbarra2021}, allows real-time tracking of the frequency shift of multiple mechanical modes at the same time. In the point-mass approximation, an adsorption event produces a modal mechanical frequency shift of the resonator:
\begin{equation}
	\label{eq:Sauerbrey}
	\frac{\Delta f_\mathrm{m}}{f_\mathrm{m}}=-\frac{\mathrm{m}}{2m_\mathrm{eff}} u(\mathbf{r_\mathrm{0}})^2
\end{equation}
that depends on the resonance frequency $f_\mathrm{m}$, on the effective mass $m_\mathrm{eff}$ of the considered mechanical mode, on the adsorbed mass $\mathrm{m}$, and on the normalized modal displacement $u(\mathbf{r_\mathrm{0}})$ at the analyte landing position $\mathbf{r_\mathrm{0}}$. In our set-up, we mounted an imaging system using a microscope objective and a fast camera (acquiring 10$^3$~frames/s) to estimate $\mathbf{r_\mathrm{0}}$ in situ in real-time when nanoparticules land on the resonator (Figure~\ref{fig:Figure1}B and Supporting Information). After optical experiments, a Scanning Electron Microscope (SEM) is employed to image the decorated resonator and increase the resolution on radial and azimuthal landing coordinates $r_\mathrm{0}$ and $\theta_0$ (Figure~\ref{fig:Figure1}C). Having in mind prior work on multimodal mechanical sensing with cantilevers \cite{Hanay2015, Malvar2016}, we identified several mechanical modes of the disk as good candidates for such purpose. These modes appear as the prominent peaks in the optomechanically measured Brownian motion spectrum of the resonator, with frequencies spanning from 100 to 600 MHz (Figure~\ref{fig:Figure1}D). Some belong to the family of Radial Breathing Modes (RBM), one is a Wine Glass Mode (M190) and two others are less classified high-order modes (M290, M490). They all mainly consist of "in-plane" vibrations of the disk, which favor their coupling to optical Whispering Gallery Modes (WGMs). In our set-up we achieved simultaneous tracking of four of these mechanical modes in real-time, with a frequency stability of 5$\times$10$^{-8}$ reached by RBM1 (Figure S1). With an effective mass of 250~pg for this mode, this leads to a minimum detectable mass of 27~ag, corresponding to a sphere of diameter 40~nm with the density of water. This is the dimension of a small virus such as Phi29 (Bacillus phage).

\section{Results and discussion}
\begin{figure}[h!]
	\includegraphics[width=8.5cm]{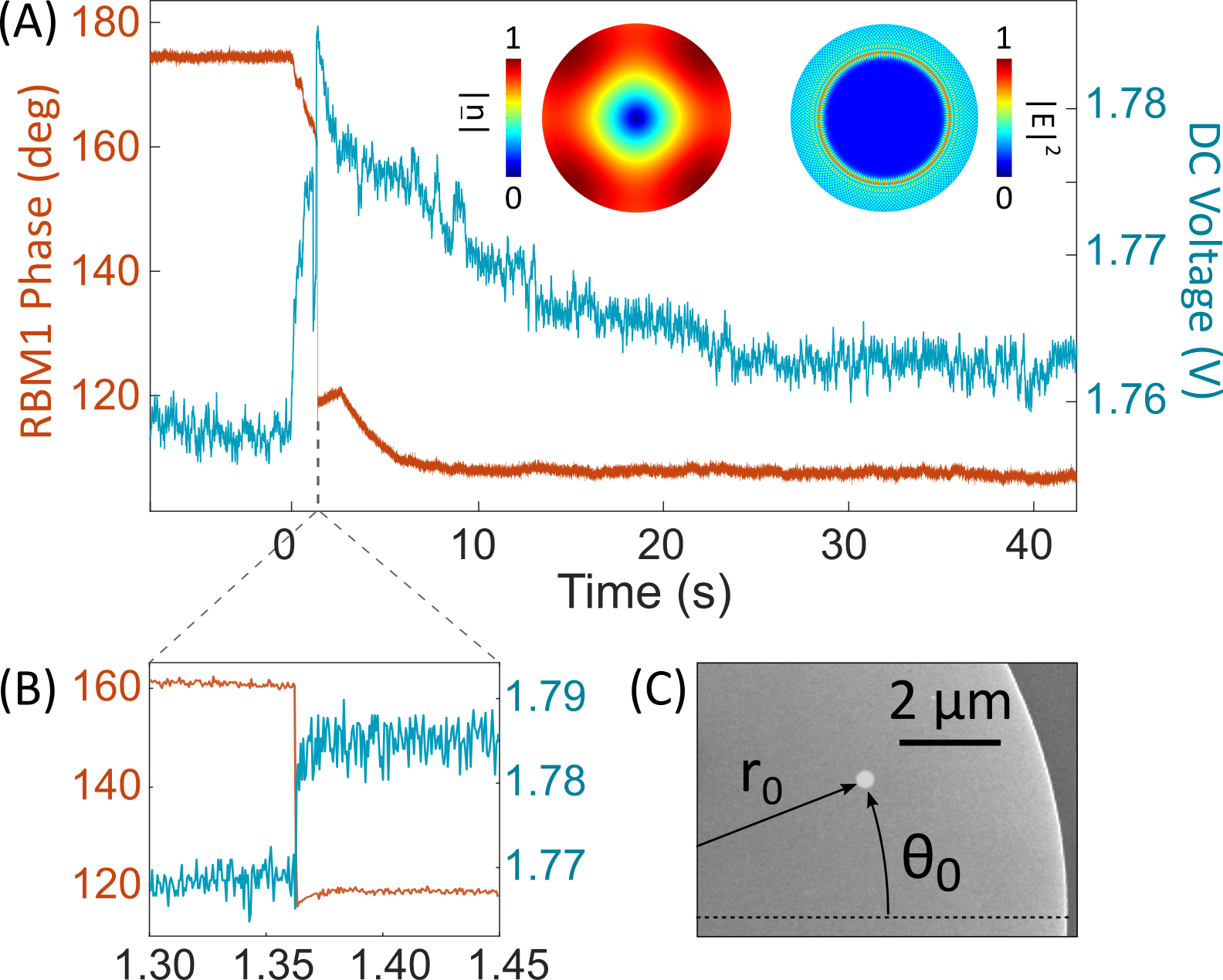}
	\caption{(A) The RF phase component of the optical output demodulated at the RBM1 resonance frequency (in red, 1~kHz bandwidth), as well as the DC optical output (blue) during a single latex nanoparticle deposition event. The displacement profile of the RBM1 and the modulus squared of the WGM electric field employed in this measurement are shown in inset. (B) A close-up on abrupt signal jumps in correspondence of the landing event. (C) The nanoparticles lands at position r$_0$=7.65~$\mu$m and $\theta_0$=26$^\circ$, measured with respect to the <100> crystalline axis.}
	\label{fig:Figure2}
\end{figure}

We first test our device by depositing latex nanoparticles on its surface. Originally in suspension in a 2-propanol solution, the nanoparticles have a nominal diameter of $300\pm{15}$~nm. They are diluted (1.35$\times$10$^9$~particles/mL) and sprayed with a piezo-ceramic nebulizer, which results in the generation of a mist surrounding the resonator. The resonator is optically probed with the input laser tuned onto the blue flank of a WGM resonance. Figure \ref{fig:Figure2}A shows time traces, during a spray, of the demodulated phase signal close to the mechanical RBM1 frequency (red), and of the DC output optical power converted into an electrical signal by a photodetector (DC voltage). By acquiring these two signals, we analyse in parallel the effect of particles deposition on both the mechanical RBM1 and optical WGM, whose mode profiles are shown in inset. Once the nebulizer is activated (t=0~s), the 2-propanol mist is generated and guided by the flux towards the resonator. The constituting droplets and molecules modify the surroundings of the resonator and waveguide, and deposit on their surfaces. Immediately after nebulization, the RF phase decreases, which is consistent with the formation of a growing liquid layer of 2-propanol on the resonator surface, red-shifting the mechanical resonance. At the same time, the device output power increases, which under our conditions is consistent with a red-shift of the WGM produced by such growing layer. A few seconds after nebulization, the output power starts in contrast to decrease, which indicates that the mist diffuses away and that the thickness of the deposited layer is progressively decreasing upon evaporation. After the end of the spray, which generally lasts around one second, this evolution goes on until a new steady-state state is reached, after few tens of seconds. The thin layer remains invisible to our optical imaging system all along, besides a blur in our camera images created by the mist. On top of this slow evolution (second timescale) associated with the sprayed mist, the landing of the nanoparticle on the disk at t$\sim$1.36~s is recognizable by a large and abrupt jump in the two signals, mechanical and optical, followed by a small and rapid transient (10 ms timescale) corresponding to the evaporation of a residual liquid enveloppe (see Figure~\ref{fig:Figure1}B). This landing event is concomitantly observed by the fast camera, which allows associating the jump to the arrival of a nanoparticle at a specific position $\mathbf{r_\mathrm{0}}$ on the disk. When required, the landing coordinates $r_\mathrm{0}$ and $\theta_0$ are finely measured in the SEM (Figure~\ref{fig:Figure2}C) after sensing experiments.

The RF phase jump of -42~deg is converted into a mechanical frequency shift of -4.1~kHz by prior calibration of the phase-frequency response (slope of -10.35~deg/kHz). This shift can be compared to analytical (Eq.~\ref{eq:Sauerbrey}) and Finite Element Method (FEM) calculations. While FEM accounts for the possible effects of particle geometry and stiffness \cite{Malvar2016, Sader2018}, the point-mass approximation of Eq.~\ref{eq:Sauerbrey} generally provides a first satisfactory estimate of the mass for our spherical particles that have moderate rigidity (detailed discussion of this approximation will be made further). From the measured frequency shift and landing position, and from the knowledge of RBM1 mode profile, we analytically deduce for the landed nanoparticle a mass of 22.4 fg. For this specific event, this comes relatively close to the mass anticipated from informations given by the commercial supplier, such as particle dimensions and latex density (14.8~fg), still with a 46\% difference between the two values. Several possible sources of deviation will be discussed along the article.\\

Simultaneous to the mechanical RF signal, the nanoparticle landing onto the disk produces an abrupt drop in the DC output optical signal. There was no observable dissipative optical effect associated to the nanoparticle landing, hence the drop relates to a red-shift of the WGM. This dispersive effect is associated to the polarizability of the particle, and a perturbative formula is often employed to model the shift \cite{Vollmer2003}:
\begin{equation}
		\label{eqn:Optical_Shift_2}
		\frac{\Delta \omega_\mathrm{opt}}{\omega_\mathrm{opt}}=-\frac{1}{2} \frac{ \alpha_\mathrm{V} \int_{V_\mathrm{np}} |E(r)|^2 dr}{\epsilon_\mathrm{0}  \int_{V_\mathrm{WGM}} \epsilon_\mathrm{r}(r) |E(r)|^2 dr}
\end{equation}
where $\alpha_\mathrm{V}$ is the particle volume polarizability in air ($\alpha_\mathrm{V}=3 \epsilon_\mathrm{0} (\epsilon_\mathrm{r}-1)$) $/(\epsilon_\mathrm{r}+2$) with $\epsilon_\mathrm{r}$ the dielectric permittivity). $|E(r)|^2$ is the modulus squared of the WGM electric field, integrated over the nanoparticle volume ($V_\mathrm{np}$) or over the optical mode volume ($V_\mathrm{WGM}$). Provided these parameters and the landing position with respect to the WGM are known, Eq.~\ref{eqn:Optical_Shift_2} enables estimating $V_\mathrm{np}$ out of the optical shift. In our experiments we deduce the optical shift from the change in the DC output signal, using a prior calibration of the thermo-optic response of the resonator  \cite{Parrain2015}. For the event reported in Figure~\ref{fig:Figure1}B, the 20~mV voltage jump corresponds to a 6.7~pm red-shift for the WGM resonant wavelength. Because the shift strongly depends on the azimuthal position of the nanoparticle with respect to the azimutal lobes of the WGM, which is not precisely known, an averaging approach brings us first to infer that the volume of the nanoparticle is of $3.09\times10^{-20}$m$^{3}$ with $\pm{55}\%$ uncertainty. If the single nanoparticle of Figure~\ref{fig:Figure2} did land on a lobe of the WGM, the lower estimation must be kept: the nanoparticle volume is evaluated to be of $1.39\times10^{-20}$m$^{3}$, which comes very close to the nominal value of $1.41\times10^{-20}$m$^{3}$ (14.4~fg) indicated by the supplier. This configuration would be consistent with a picture where optical gradient forces would favor the deposition of dielectric nanoparticles on the lobes of the mode. This may however not always be the case, and such residual uncertainty in the estimation of the particle volume from a single optical signal illustrates the interest of acquiring multiple signals at the same time. In our work, multiple information is provided by dual optical mechanical interactions of the analyte with the resonator.

\begin{figure}[h!]
	\includegraphics[width=8.5cm]{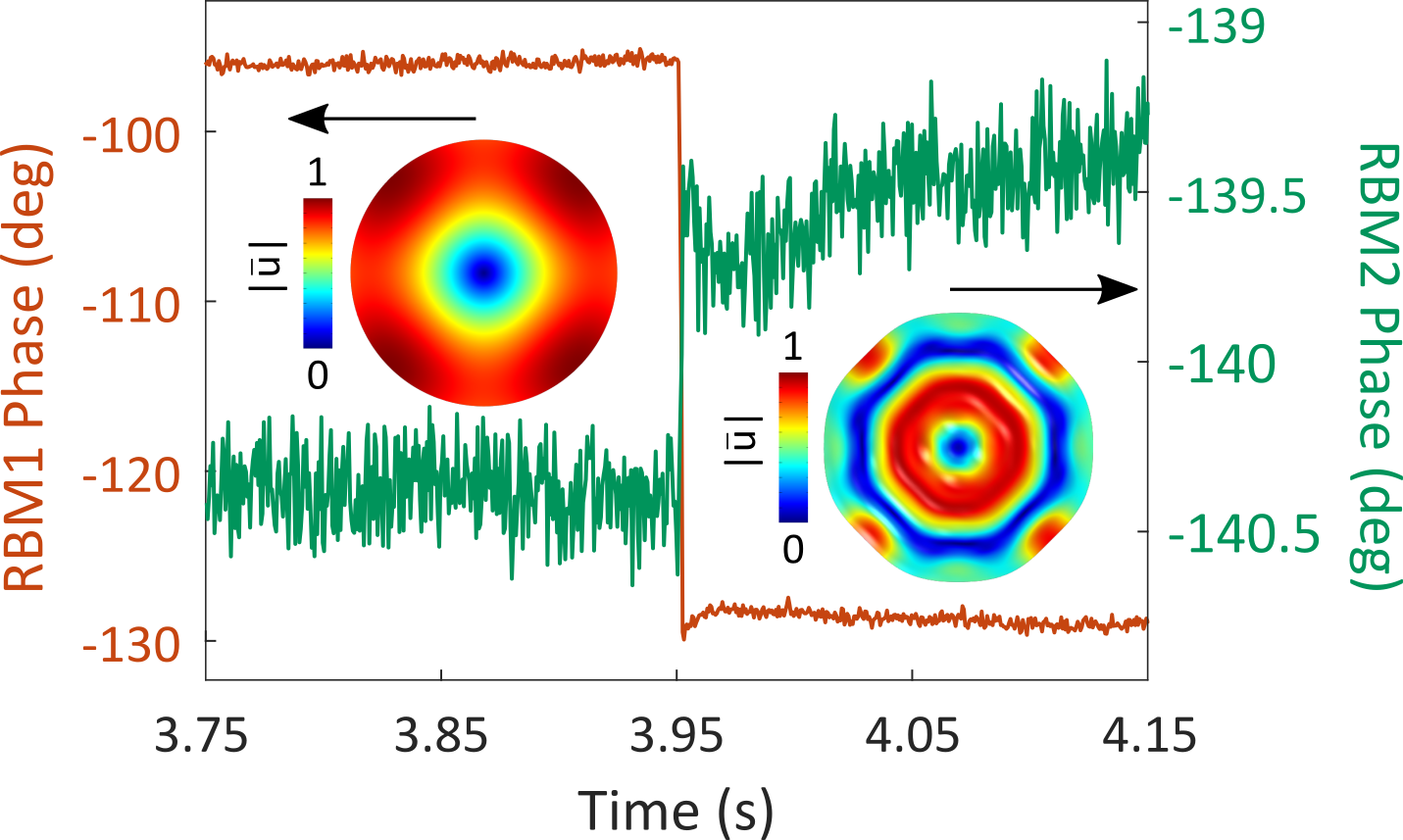}
	\caption{The phase of the demodulated output signal at the RBM1 (in red) and RBM2 (in green) resonance frequency during the landing of a cluster of three latex nanoparticles. The displacement mode profile of the two mechanical modes is reported in inset.}
	\label{fig:Figure3}
\end{figure}

In the same spirit, we employ next our device to perform multimode mechanical weighing of latex nanoparticles by simultaneously tracking both RBM1 and RBM2. The phase signals for both mechanical modes are reported in Figure~\ref{fig:Figure3}, together with their mode profiles in inset, as a cluster of three nanoparticles is landing onto the disk. Because of a comparatively smaller mechanical quality factor and optomechanical coupling, the signal from RBM2 is more noisy and the phase jump smaller. The WGM optical density at the particles landing position is much smaller than in the prior case, leading to a negligible optical resonance shift for the present event. As evidenced by subsequent SEM imaging, the three landed particles are touching one another, forming a triangle at the surface of the disk that ensures equivalent contact of each particle with the disk surface after landing. This condition guarantees that the disk vibration is coupled equivalently to each nanoparticle. Within the point-mass approximation of Eq.~\ref{eq:Sauerbrey}, the mass of the cluster deduced from the signals of RBM1 and RBM2 is respectively of 49.8~fg and 26.1~fg, i.e. $12\%$ more and $41\%$ less than the value of 44.4 fg expected from the supplier parameters. A FEM analysis shows that the mode profile of RBM2 is distorted when the shape of the pedestal sustaining the disk is non-cylindrical, thereby affecting the estimation of the landing mass from Eq. \ref{eq:Sauerbrey}. This effect is sizable when the landing position is close to the pedestal, and less pronounced close to the disk periphery. In our experiments, it induces a larger mass uncertainty when using RBM2, while the modal distortion is far less pronounced for RBM1, providing more accurate mass estimation. Modes M190, 290 and 490 also present small modal distortion, and might be used for sensing in future experiments.

\begin{figure}
	\includegraphics[width=8.5cm]{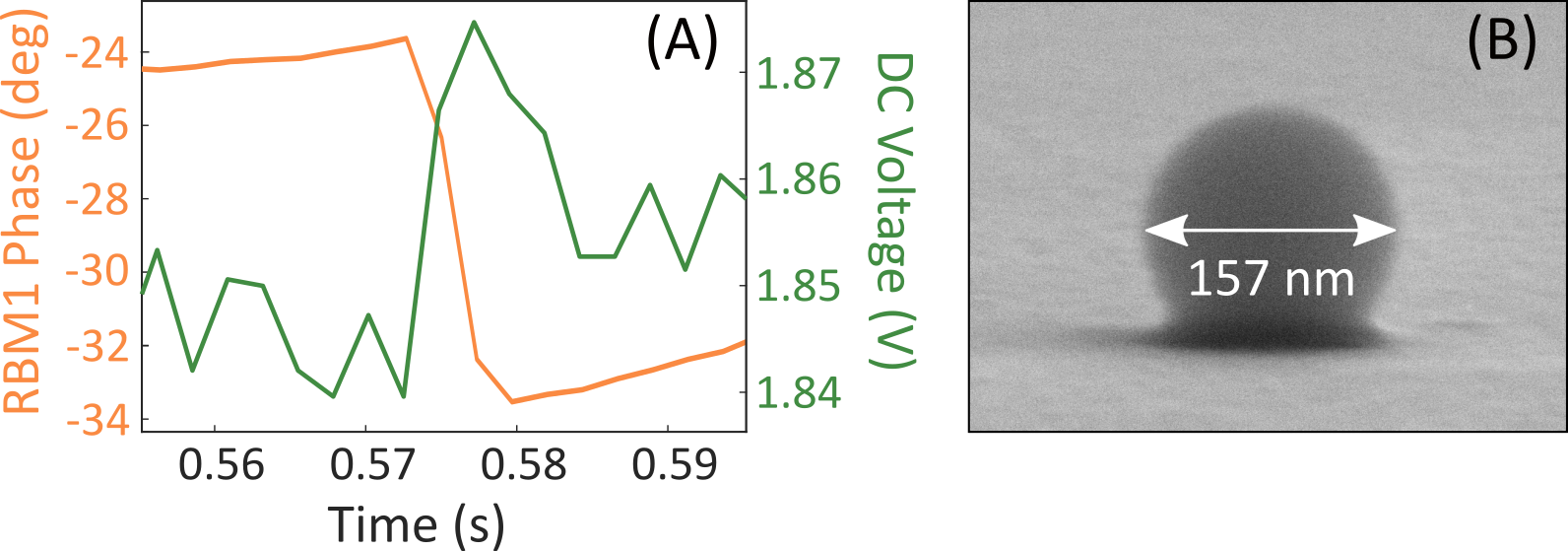}
	\caption{(A) The DC output power transduced in voltage (green) and the phase of its RF component demodulated at the RBM1 resonance frequency (orange) during a single silica nanoparticle deposition event. (B) A SEM side view of the 157~nm diameter nanoparticle at its landing position.}
	\label{fig:Figure4}
\end{figure}
We further test the sensitivity of our device by performing sensing experiments with silica nanoparticles of 150~nm nominal diameter, 8 times smaller in volume. This is about the dimension of a large virus such as SARS-CoV-2. In Figure~\ref{fig:Figure4}A, the RBM1 phase signal and the DC optical output voltage are tracked in time during the landing of a single silica nanoparticle on the disk surface. A 100~Hz measurement bandwidth was employed, which was shown to provide the best sensitivity (Supporting Information). While the mechanical phase jump is still substantially larger than the noise, the change in DC voltage is faint and approaches the noise floor. Starting from the the RBM1 phase jump, we employ again Eq. \ref{eq:Sauerbrey} and deduce a nanoparticle mass  of 5.8~fg, which comes 50\% above the value derived from the supplier nominal informations. As visible in Figure~\ref{fig:Figure4}B, a small amount of soft matter seems to accompany the particle and to be trapped at its interface with the disk. It may be responsible for a residual additional mass in the after-landing state, which is yet another source of deviation.

\begin{figure}[h!]
	\includegraphics[width=7cm]{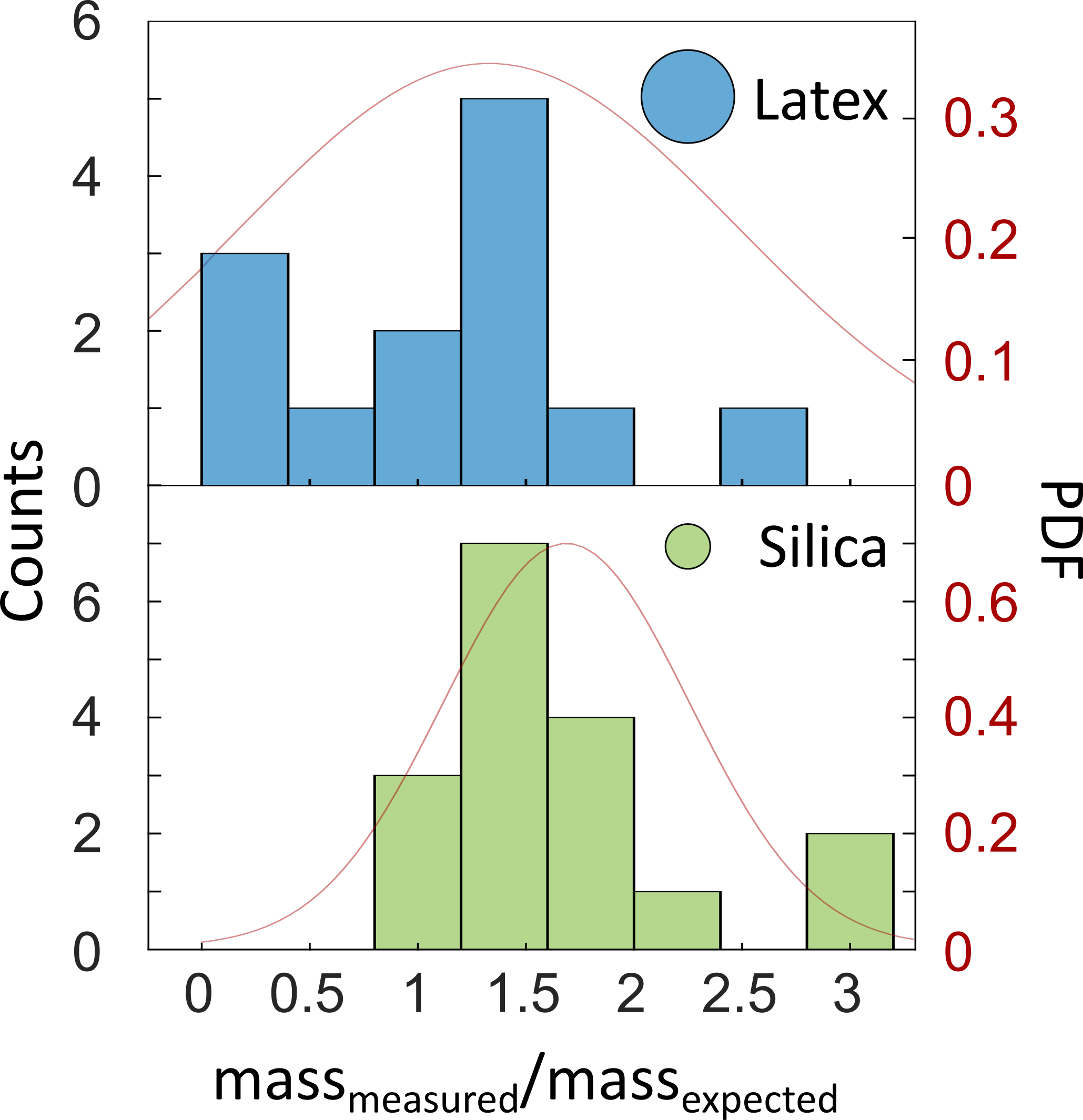}
	\caption{Histogram of latex (blue) and silica (green) nanoparticle masses measured experimentally.  Each detection event was normalized to an expected mass (see text). In case of small clusters (2 or 3 particles), the measured mass was divided by the number of nanoparticles.}
	\label{fig:Histogram}
\end{figure}

To obtain a global viewpoint on the precision of our optomechanical approach for weighing a landing nanoparticle, we analyse a larger set of data. The histograms in Figure~\ref{fig:Histogram} gathers latex (blue) and silica (green) nanoparticle masses measured by our optomechanical sensor. Each measured mass is normalized by an expected mass. For silica particles, we measured the exact dimensions of each landed particle with our calibrated SEM, and employed a referenced value for silica density to obtain the expected mass. For latex particles, we chose instead to normalize by the nominal mass corresponding to dimensions and density given by the supplier. Indeed, once extracted from the liquid solution, we observed that latex particle dimensions evolve on a minute timescale after landing, making mass estimation trough the SEM poorly reliable, in contrast to silica. In the histograms, the column width corresponds to our experimental error in evaluating the mechanical frequency shift in detection events. This evaluation is impacted by the above-mentioned mist and nanometric liquid layer forming during the spray, which modify optical and mechanical frequencies through thermal effects, hence affecting the phase-frequency slope of our RF signal. A proper characterization of the device and the application of the actuation/detection model of ref.~\citenum{Sbarra2021} enable decoupling this contribution from that of the nanoparticle mass, with a residual error of $\pm$20\%.
A few conclusions can be drawn out of these histograms. First, optomechanical weighing seems to consistently provide a mass that is within a factor 2 of the expectation. Despite this overall satisfactory agreement, the histograms are centered close to 1.5 and come with a sizable dispersion. Sources of dispersion were already discussed above: inaccurate a priori knowledge of the deposited particles, deformation by the pedestal of the employed mechanical modes, and error in estimating mechanical shifts. On top come systematic deviations. The first is the presence of a residual liquid meniscus trapped between the nanoparticle and the disk, whose relative mass contribution is more important for smallest particles, and is anticipated to dominate the deviation for our silica particles. This effect is less relevant for our latex particles, whose volume sits a decade above. The second source of deviation though is the mechanical coupling between the disk and nanoparticle vibrational modes, recently addressed in ref.~\citenum{Gil-Santos2020}. A spherical nanoparticle attached to a surface displays a flexural mode whose oscillation frequency depends on its dimensions, elasticity, density and contact area with the surface. FEM calculations show that the in-plane vibrations of the disk can efficiently excite such flexural mode of our latex nanoparticles, while the effect is negligible for our silica particles. In this "mechanical coupling regime", the point-mass approximation becomes inadequate, leading to an overestimation of the adsorbed mass (see Supporting Information). With the contact radius of 61~nm we measure in the SEM, the numerically calculated shift for the event of Figure~\ref{fig:Figure2} is for example 1.5 times larger than the point-mass approximation result, explaining the off-centering of the histogram for latex nanoparticles. Conversely, if the residual liquid effect can be neglected for our latex particles, the deviated shifts enable analyzing the nanoparticle elasticity, provided its density is known. From our set of measurements, and with an average contact radius of $56\pm5$nm, we obtain this way a material Young modulus of $2.7\pm0.6$GPa, which is consistent with a referenced value of 3 GPa for latex.

\section{Conclusions}
In conclusion, we demonstrate that multimode weighing of individual nanoparticles can be achieved by an optomechanical sensor. This approach enables accessing multiple mechanical and optical informations in real-time, which may lead to the analysis of several physical properties of the analyte at once, beyond its mere mass. The quantitative modeling of our sensing experiments reveals that a fine control and understanding of every aspect is important, from the device physics and engineering to the analyte pre-treatment and guiding to the sensor. The current device, with a large capture area that approaches state-of-the-art focusing capabilities for Electro-Spray Ionization methods, is well suited for future investigation of biological particles. This is notably the case of a large range of viruses with masses in the tens of MDa range, which could be adressed by such multiphysics approach.

\begin{suppinfo}
Measurements of the frequency Allan variance of four mechanical modes, for a simultaneous all-optical actuation/detection, can be found in the Supporting Information. Examples of video frames acquired by the fast camera during a nanoparticle landing event, and of SEM figures of silica and latex nanoparticles deposited on the optomechanical resonator, are also shown. Eventually, we discuss the regime of mechanical coupling between the nanoparticle and the disk.
\end{suppinfo}

\begin{acknowledgement}
The authors acknowledge support from the European Commission through the VIRUSCAN (731868) FET-open and NOMLI (770933) ERC projects, and from the Agence Nationale de la Recherche through the QuaSeRT Quantera project. They also thank Dimitris Papanastasiou, S{\'{e}}bastien Hentz, Adrien Daerr, Christophe Masselon and the Bionanomechanics team of IMN-CSIC for discussions.
\end{acknowledgement}

\bibliography{library}

\providecommand{\latin}[1]{#1}
\makeatletter
\providecommand{\doi}
  {\begingroup\let\do\@makeother\dospecials
  \catcode`\{=1 \catcode`\}=2 \doi@aux}
\providecommand{\doi@aux}[1]{\endgroup\texttt{#1}}
\makeatother
\providecommand*\mcitethebibliography{\thebibliography}
\csname @ifundefined\endcsname{endmcitethebibliography}
  {\let\endmcitethebibliography\endthebibliography}{}
\begin{mcitethebibliography}{28}
\providecommand*\natexlab[1]{#1}
\providecommand*\mciteSetBstSublistMode[1]{}
\providecommand*\mciteSetBstMaxWidthForm[2]{}
\providecommand*\mciteBstWouldAddEndPuncttrue
  {\def\EndOfBibitem{\unskip.}}
\providecommand*\mciteBstWouldAddEndPunctfalse
  {\let\EndOfBibitem\relax}
\providecommand*\mciteSetBstMidEndSepPunct[3]{}
\providecommand*\mciteSetBstSublistLabelBeginEnd[3]{}
\providecommand*\EndOfBibitem{}
\mciteSetBstSublistMode{f}
\mciteSetBstMaxWidthForm{subitem}{(\alph{mcitesubitemcount})}
\mciteSetBstSublistLabelBeginEnd
  {\mcitemaxwidthsubitemform\space}
  {\relax}
  {\relax}

\bibitem[Chaste \latin{et~al.}(2012)Chaste, Eichler, Moser, Ceballos, Rurali,
  and Bachtold]{Chaste2012}
Chaste,~J.; Eichler,~A.; Moser,~J.; Ceballos,~G.; Rurali,~R.; Bachtold,~A. {A
  nanomechanical mass sensor with yoctogram resolution}. \emph{Nature
  Nanotechnology} \textbf{2012}, \emph{7}, 301--304\relax
\mciteBstWouldAddEndPuncttrue
\mciteSetBstMidEndSepPunct{\mcitedefaultmidpunct}
{\mcitedefaultendpunct}{\mcitedefaultseppunct}\relax
\EndOfBibitem
\bibitem[Hanay \latin{et~al.}(2012)Hanay, Kelber, Naik, Chi, Hentz, Bullard,
  Colinet, Duraffourg, and Roukes]{Hanay2012}
Hanay,~M.~S.; Kelber,~S.; Naik,~A.~K.; Chi,~D.; Hentz,~S.; Bullard,~E.~C.;
  Colinet,~E.; Duraffourg,~L.; Roukes,~M.~L. {Single-protein nanomechanical
  mass spectrometry in real time}. \emph{Nature Nanotechnology} \textbf{2012},
  \emph{7}, 602\relax
\mciteBstWouldAddEndPuncttrue
\mciteSetBstMidEndSepPunct{\mcitedefaultmidpunct}
{\mcitedefaultendpunct}{\mcitedefaultseppunct}\relax
\EndOfBibitem
\bibitem[Dominguez-Medina \latin{et~al.}(2018)Dominguez-Medina, Fostner,
  Defoort, Sansa, Stark, Halim, Vernhes, Gely, Jourdan, Alava, Boulanger,
  Masselon, and Hentz]{HentzCapsid2018}
Dominguez-Medina,~S.; Fostner,~S.; Defoort,~M.; Sansa,~M.; Stark,~A.-K.;
  Halim,~M.~A.; Vernhes,~E.; Gely,~M.; Jourdan,~G.; Alava,~T.; Boulanger,~P.;
  Masselon,~C.; Hentz,~S. {Neutral mass spectrometry of virus capsids above 100
  megadaltons with nanomechanical resonators}. \emph{Science} \textbf{2018},
  \emph{362}, 918--922\relax
\mciteBstWouldAddEndPuncttrue
\mciteSetBstMidEndSepPunct{\mcitedefaultmidpunct}
{\mcitedefaultendpunct}{\mcitedefaultseppunct}\relax
\EndOfBibitem
\bibitem[Sauer \latin{et~al.}(2017)Sauer, Diao, Westwood-Bachman, Freeman, and
  Hiebert]{Sauer2017}
Sauer,~V.~T.; Diao,~Z.; Westwood-Bachman,~J.~N.; Freeman,~M.~R.; Hiebert,~W.~K.
  {Single laser modulated drive and detection of a nano-optomechanical
  cantilever}. \emph{AIP Advances} \textbf{2017}, \emph{7}\relax
\mciteBstWouldAddEndPuncttrue
\mciteSetBstMidEndSepPunct{\mcitedefaultmidpunct}
{\mcitedefaultendpunct}{\mcitedefaultseppunct}\relax
\EndOfBibitem
\bibitem[Allain \latin{et~al.}(2020)Allain, Schwab, Mismer, Gely, Mairiaux,
  Hermouet, Walter, Leo, Hentz, Faucher, Jourdan, Legrand, and
  Favero]{Allain2020}
Allain,~P.~E.; Schwab,~L.; Mismer,~C.; Gely,~M.; Mairiaux,~E.; Hermouet,~M.;
  Walter,~B.; Leo,~G.; Hentz,~S.; Faucher,~M.; Jourdan,~G.; Legrand,~B.;
  Favero,~I. {Optomechanical resonating probe for very high frequency sensing
  of atomic forces}. \emph{Nanoscale} \textbf{2020}, \emph{12}, 2939\relax
\mciteBstWouldAddEndPuncttrue
\mciteSetBstMidEndSepPunct{\mcitedefaultmidpunct}
{\mcitedefaultendpunct}{\mcitedefaultseppunct}\relax
\EndOfBibitem
\bibitem[Guha \latin{et~al.}(2020)Guha, Allain, Lema{\^{i}}tre, Leo, and
  Favero]{Guha2020}
Guha,~B.; Allain,~P.~E.; Lema{\^{i}}tre,~A.; Leo,~G.; Favero,~I. {Force Sensing
  with an Optomechanical Self-Oscillator}. \emph{Physical Review Applied}
  \textbf{2020}, \emph{14}, 024079\relax
\mciteBstWouldAddEndPuncttrue
\mciteSetBstMidEndSepPunct{\mcitedefaultmidpunct}
{\mcitedefaultendpunct}{\mcitedefaultseppunct}\relax
\EndOfBibitem
\bibitem[Favero and Karrai(2009)Favero, and Karrai]{Favero2009}
Favero,~I.; Karrai,~K. {Optomechanics of deformable optical cavities}.
  \emph{Nature Photonics} \textbf{2009}, \emph{3}, 201--205\relax
\mciteBstWouldAddEndPuncttrue
\mciteSetBstMidEndSepPunct{\mcitedefaultmidpunct}
{\mcitedefaultendpunct}{\mcitedefaultseppunct}\relax
\EndOfBibitem
\bibitem[Aspelmeyer \latin{et~al.}(2014)Aspelmeyer, Kippenberg, and
  Marquardt]{Aspelmeyer2014}
Aspelmeyer,~M.; Kippenberg,~T.~J.; Marquardt,~F. {Cavity optomechanics}.
  \emph{Reviews of Modern Physics} \textbf{2014}, \emph{86}, 1391--1452\relax
\mciteBstWouldAddEndPuncttrue
\mciteSetBstMidEndSepPunct{\mcitedefaultmidpunct}
{\mcitedefaultendpunct}{\mcitedefaultseppunct}\relax
\EndOfBibitem
\bibitem[Liu \latin{et~al.}(2013)Liu, Alaie, Leseman, and
  Hossein-Zadeh]{Liu2013}
Liu,~F.; Alaie,~S.; Leseman,~Z.~C.; Hossein-Zadeh,~M. {Sub-pg mass sensing and
  measurement with an optomechanical oscillator}. \emph{Optics Express}
  \textbf{2013}, \emph{21}, 19555\relax
\mciteBstWouldAddEndPuncttrue
\mciteSetBstMidEndSepPunct{\mcitedefaultmidpunct}
{\mcitedefaultendpunct}{\mcitedefaultseppunct}\relax
\EndOfBibitem
\bibitem[Yu \latin{et~al.}(2016)Yu, Jiang, Lin, and Lu]{Yu2016}
Yu,~W.; Jiang,~W.~C.; Lin,~Q.; Lu,~T. {Cavity optomechanical spring sensing of
  single molecules}. \emph{Nature Communications} \textbf{2016}, \emph{7},
  12311\relax
\mciteBstWouldAddEndPuncttrue
\mciteSetBstMidEndSepPunct{\mcitedefaultmidpunct}
{\mcitedefaultendpunct}{\mcitedefaultseppunct}\relax
\EndOfBibitem
\bibitem[Venkatasubramanian \latin{et~al.}(2016)Venkatasubramanian, Sauer, Roy,
  Xia, Wishart, and Hiebert]{Venkatasubramanian2016}
Venkatasubramanian,~A.; Sauer,~V.~T.; Roy,~S.~K.; Xia,~M.; Wishart,~D.~S.;
  Hiebert,~W.~K. {Nano-Optomechanical Systems for Gas Chromatography}.
  \emph{Nano Letters} \textbf{2016}, \emph{16}, 6975--6981\relax
\mciteBstWouldAddEndPuncttrue
\mciteSetBstMidEndSepPunct{\mcitedefaultmidpunct}
{\mcitedefaultendpunct}{\mcitedefaultseppunct}\relax
\EndOfBibitem
\bibitem[Maksymowych \latin{et~al.}(2019)Maksymowych, Westwood-Bachman,
  Venkatasubramanian, and Hiebert]{Maksymowych2019}
Maksymowych,~M.~P.; Westwood-Bachman,~J.~N.; Venkatasubramanian,~A.;
  Hiebert,~W.~K. {Optomechanical spring enhanced mass sensing}. \emph{Applied
  Physics Letters} \textbf{2019}, \emph{115}\relax
\mciteBstWouldAddEndPuncttrue
\mciteSetBstMidEndSepPunct{\mcitedefaultmidpunct}
{\mcitedefaultendpunct}{\mcitedefaultseppunct}\relax
\EndOfBibitem
\bibitem[Gil-Santos \latin{et~al.}(2020)Gil-Santos, Ruz, Malvar, Favero,
  Lema{\^{i}}tre, Kosaka, Garc{\'{i}}a-L{\'{o}}pez, Calleja, and
  Tamayo]{Gil-Santos2020}
Gil-Santos,~E.; Ruz,~J.~J.; Malvar,~O.; Favero,~I.; Lema{\^{i}}tre,~A.;
  Kosaka,~P.~M.; Garc{\'{i}}a-L{\'{o}}pez,~S.; Calleja,~M.; Tamayo,~J.
  {Optomechanical detection of vibration modes of a single bacterium}.
  \emph{Nature Nanotechnology} \textbf{2020}, \emph{15}, 469--474\relax
\mciteBstWouldAddEndPuncttrue
\mciteSetBstMidEndSepPunct{\mcitedefaultmidpunct}
{\mcitedefaultendpunct}{\mcitedefaultseppunct}\relax
\EndOfBibitem
\bibitem[Sansa \latin{et~al.}(2020)Sansa, Defoort, Brenac, Hermouet, Banniard,
  Fafin, Gely, Masselon, Favero, Jourdan, and Hentz]{Sansa2020a}
Sansa,~M.; Defoort,~M.; Brenac,~A.; Hermouet,~M.; Banniard,~L.; Fafin,~A.;
  Gely,~M.; Masselon,~C.; Favero,~I.; Jourdan,~G.; Hentz,~S. {Optomechanical
  mass spectrometry}. \emph{Nature Communications} \textbf{2020}, \emph{11},
  3781\relax
\mciteBstWouldAddEndPuncttrue
\mciteSetBstMidEndSepPunct{\mcitedefaultmidpunct}
{\mcitedefaultendpunct}{\mcitedefaultseppunct}\relax
\EndOfBibitem
\bibitem[Ding \latin{et~al.}(2010)Ding, Baker, Senellart, Lemaitre, Ducci, Leo,
  and Favero]{Ding2010}
Ding,~L.; Baker,~C.; Senellart,~P.; Lemaitre,~A.; Ducci,~S.; Leo,~G.;
  Favero,~I. {High frequency GaAs nano-optomechanical disk resonator}.
  \emph{Physical Review Letters} \textbf{2010}, \emph{105}, 1--4\relax
\mciteBstWouldAddEndPuncttrue
\mciteSetBstMidEndSepPunct{\mcitedefaultmidpunct}
{\mcitedefaultendpunct}{\mcitedefaultseppunct}\relax
\EndOfBibitem
\bibitem[Eichenfield \latin{et~al.}(2009)Eichenfield, Chan, Camacho, Vahala,
  and Painter]{Eichenfield2009}
Eichenfield,~M.; Chan,~J.; Camacho,~R.~M.; Vahala,~K.~J.; Painter,~O.
  {Optomechanical crystals}. \emph{Nature} \textbf{2009}, \emph{462},
  78--82\relax
\mciteBstWouldAddEndPuncttrue
\mciteSetBstMidEndSepPunct{\mcitedefaultmidpunct}
{\mcitedefaultendpunct}{\mcitedefaultseppunct}\relax
\EndOfBibitem
\bibitem[Baker \latin{et~al.}(2014)Baker, Hease, Nguyen, Andronico, Ducci, Leo,
  and Favero]{Baker2014b}
Baker,~C.; Hease,~W.; Nguyen,~D.-T.; Andronico,~A.; Ducci,~S.; Leo,~G.;
  Favero,~I. {Photoelastic coupling in gallium arsenide optomechanical disk
  resonators}. \emph{Optics Express} \textbf{2014}, \emph{22}, 14072\relax
\mciteBstWouldAddEndPuncttrue
\mciteSetBstMidEndSepPunct{\mcitedefaultmidpunct}
{\mcitedefaultendpunct}{\mcitedefaultseppunct}\relax
\EndOfBibitem
\bibitem[Sbarra \latin{et~al.}(2021)Sbarra, Allain, Lema{\^{i}}tre, and
  Favero]{Sbarra2021}
Sbarra,~S.; Allain,~P.~E.; Lema{\^{i}}tre,~A.; Favero,~I. {A multiphysics model
  for high frequency optomechanical sensors optically actuated and detected in
  the oscillating mode}. \textbf{2021}, \emph{086111}, 1--6\relax
\mciteBstWouldAddEndPuncttrue
\mciteSetBstMidEndSepPunct{\mcitedefaultmidpunct}
{\mcitedefaultendpunct}{\mcitedefaultseppunct}\relax
\EndOfBibitem
\bibitem[Gil-Santos \latin{et~al.}(2015)Gil-Santos, Baker, Nguyen, Hease,
  Gomez, Lema{\^{i}}tre, Ducci, Leo, and Favero]{Gil-Santos2015}
Gil-Santos,~E.; Baker,~C.; Nguyen,~D.~T.; Hease,~W.; Gomez,~C.;
  Lema{\^{i}}tre,~A.; Ducci,~S.; Leo,~G.; Favero,~I. {High-frequency
  nano-optomechanical disk resonators in liquids}. \emph{Nature Nanotechnology}
  \textbf{2015}, \emph{10}, 810--816\relax
\mciteBstWouldAddEndPuncttrue
\mciteSetBstMidEndSepPunct{\mcitedefaultmidpunct}
{\mcitedefaultendpunct}{\mcitedefaultseppunct}\relax
\EndOfBibitem
\bibitem[Fong \latin{et~al.}(2015)Fong, Poot, and Tang]{Fong2015}
Fong,~K.~Y.; Poot,~M.; Tang,~H.~X. {Nano-Optomechanical Resonators in
  Microfluidics}. \emph{Nano Letters} \textbf{2015}, \emph{15},
  6116--6120\relax
\mciteBstWouldAddEndPuncttrue
\mciteSetBstMidEndSepPunct{\mcitedefaultmidpunct}
{\mcitedefaultendpunct}{\mcitedefaultseppunct}\relax
\EndOfBibitem
\bibitem[Hermouet(2019)]{Hermouet2019}
Hermouet,~M. {Microdisques optom{\'{e}}caniques r{\'{e}}sonants en silicium
  pour la d{\'{e}}tection biologique en milieu liquide Optomechanical silicon
  microdisk resonators for biosensing in liquid}. Ph.D.\ thesis, 2019\relax
\mciteBstWouldAddEndPuncttrue
\mciteSetBstMidEndSepPunct{\mcitedefaultmidpunct}
{\mcitedefaultendpunct}{\mcitedefaultseppunct}\relax
\EndOfBibitem
\bibitem[Baker \latin{et~al.}(2011)Baker, Belacel, Andronico, Senellart,
  Lema{\^{i}}tre, Galopin, Ducci, Leo, and Favero]{BakerWaveguide2011}
Baker,~C.; Belacel,~C.; Andronico,~A.; Senellart,~P.; Lema{\^{i}}tre,~A.;
  Galopin,~E.; Ducci,~S.; Leo,~G.; Favero,~I. {Critical optical coupling
  between a GaAs disk and a nanowaveguide suspended on the chip}. \emph{Applied
  Physics Letters} \textbf{2011}, \emph{99}, 151117\relax
\mciteBstWouldAddEndPuncttrue
\mciteSetBstMidEndSepPunct{\mcitedefaultmidpunct}
{\mcitedefaultendpunct}{\mcitedefaultseppunct}\relax
\EndOfBibitem
\bibitem[Hanay \latin{et~al.}(2015)Hanay, Kelber, O'Connell, Mulvaney, Sader,
  and Roukes]{Hanay2015}
Hanay,~M.~S.; Kelber,~S.~I.; O'Connell,~C.~D.; Mulvaney,~P.; Sader,~J.~E.;
  Roukes,~M.~L. {Inertial imaging with nanomechanical systems}. \emph{Nature
  Nanotechnology} \textbf{2015}, \emph{10}, 339--344\relax
\mciteBstWouldAddEndPuncttrue
\mciteSetBstMidEndSepPunct{\mcitedefaultmidpunct}
{\mcitedefaultendpunct}{\mcitedefaultseppunct}\relax
\EndOfBibitem
\bibitem[Malvar \latin{et~al.}(2016)Malvar, Ruz, Kosaka, Dom{\'{i}}nguez,
  Gil-Santos, Calleja, and Tamayo]{Malvar2016}
Malvar,~O.; Ruz,~J.~J.; Kosaka,~P.~M.; Dom{\'{i}}nguez,~C.~M.; Gil-Santos,~E.;
  Calleja,~M.; Tamayo,~J. {Mass and stiffness spectrometry of nanoparticles and
  whole intact bacteria by multimode nanomechanical resonators}. \emph{Nature
  Communications} \textbf{2016}, \emph{7}, 13452\relax
\mciteBstWouldAddEndPuncttrue
\mciteSetBstMidEndSepPunct{\mcitedefaultmidpunct}
{\mcitedefaultendpunct}{\mcitedefaultseppunct}\relax
\EndOfBibitem
\bibitem[Sader \latin{et~al.}(2018)Sader, Hanay, Neumann, and
  Roukes]{Sader2018}
Sader,~J.~E.; Hanay,~M.~S.; Neumann,~A.~P.; Roukes,~M.~L. {Mass Spectrometry
  Using Nanomechanical Systems: Beyond the Point-Mass Approximation}.
  \emph{Nano Letters} \textbf{2018}, \emph{18}, 1608--1614\relax
\mciteBstWouldAddEndPuncttrue
\mciteSetBstMidEndSepPunct{\mcitedefaultmidpunct}
{\mcitedefaultendpunct}{\mcitedefaultseppunct}\relax
\EndOfBibitem
\bibitem[Arnold \latin{et~al.}(2003)Arnold, Khoshsima, Teraoka, Holler, and
  Vollmer]{Vollmer2003}
Arnold,~S.; Khoshsima,~M.; Teraoka,~I.; Holler,~S.; Vollmer,~F. {Shift of
  whispering gallery modes in microspheres by protein adsorption}. \emph{Opt.
  Lett.} \textbf{2003}, \emph{28}, 272--274\relax
\mciteBstWouldAddEndPuncttrue
\mciteSetBstMidEndSepPunct{\mcitedefaultmidpunct}
{\mcitedefaultendpunct}{\mcitedefaultseppunct}\relax
\EndOfBibitem
\bibitem[Parrain \latin{et~al.}(2015)Parrain, Baker, Wang, Guha, Santos,
  Lemaitre, Senellart, Leo, Ducci, and Favero]{Parrain2015}
Parrain,~D.; Baker,~C.; Wang,~G.; Guha,~B.; Santos,~E.~G.; Lemaitre,~A.;
  Senellart,~P.; Leo,~G.; Ducci,~S.; Favero,~I. {Origin of optical losses in
  gallium arsenide disk whispering gallery resonators}. \emph{Optics Express}
  \textbf{2015}, \emph{23}, 19656\relax
\mciteBstWouldAddEndPuncttrue
\mciteSetBstMidEndSepPunct{\mcitedefaultmidpunct}
{\mcitedefaultendpunct}{\mcitedefaultseppunct}\relax
\EndOfBibitem
\end{mcitethebibliography}
\newpage
\onecolumn
\section{Supporting Information}

\begin{figure*}
		\includegraphics[width=0.8\textwidth]{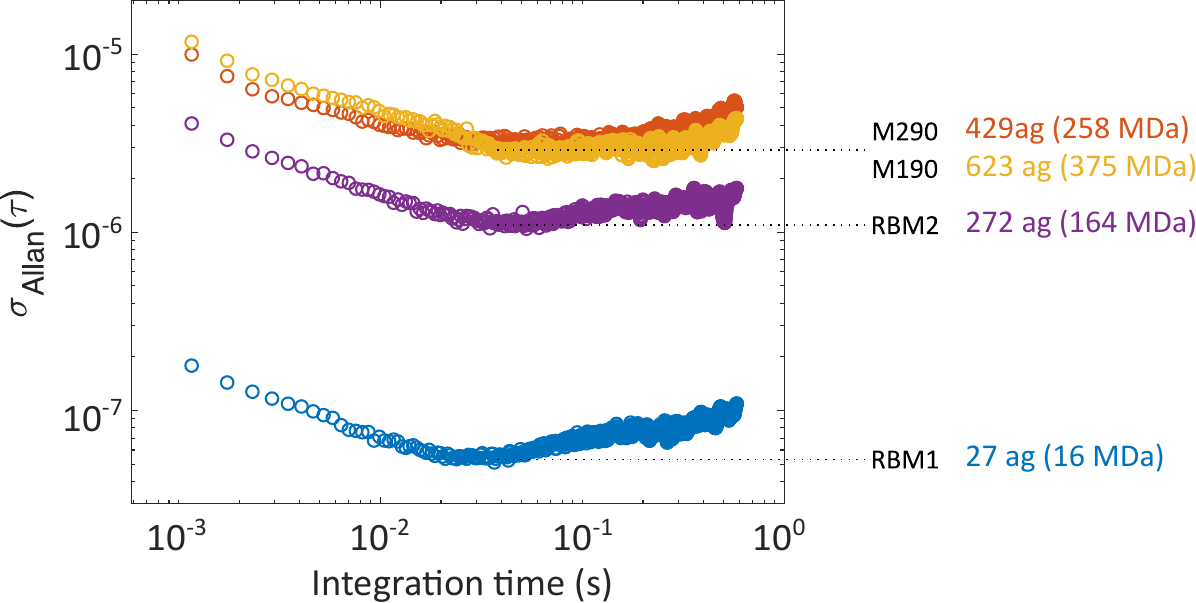}
		\caption{Frequency stability of four mechanical modes of the disk sensor. Allan variances for different integration times, in a configuration of simultaneous all-optical actuation/detection of the four modes. A laser modulation corresponding to 6\% of the injected power was chosen for each mode. The minimum detectable mass deduced from these stability data is reported on the right.}
	\end{figure*}
	
		\begin{figure*}
		\includegraphics[width=0.8\textwidth]{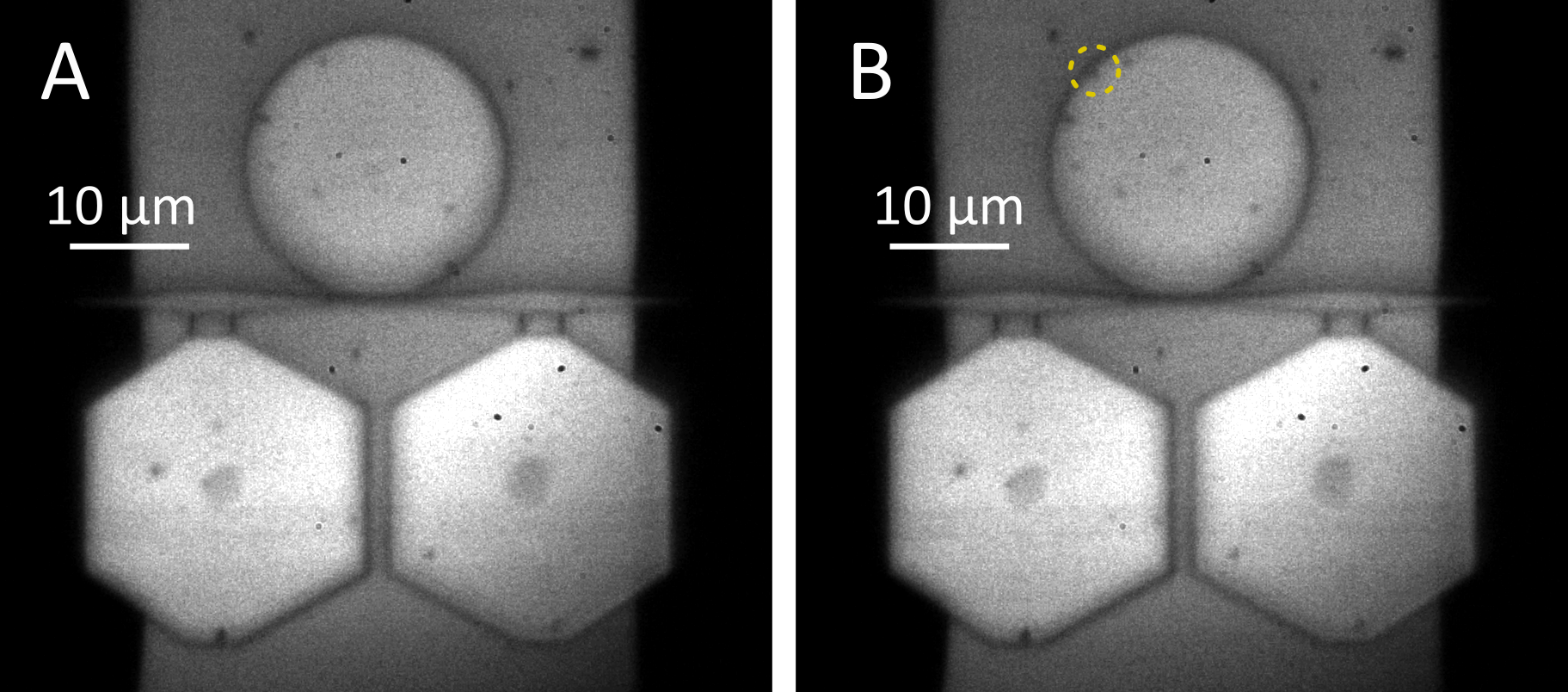}
		\caption{Two consecutive frames of fast camera, acquired during a single nanoparticle landing event occurring at the disk border (1~kHz acquisition rate). The landed nanoparticle is indicated with a yellow circle. Other particles, deposited during previous sprays, are also visible.}
	\end{figure*}

			\begin{figure*}
		\includegraphics[width=0.7\textwidth]{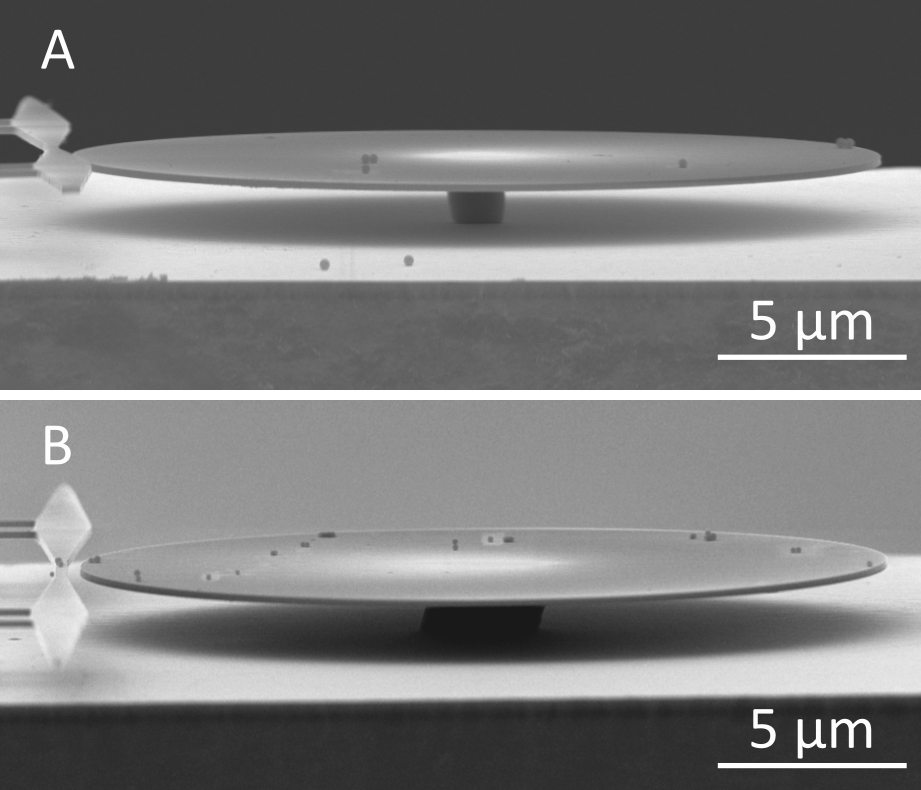}
		\caption{SEM images of latex (A) and silica (B) nanoparticles deposited over two distinct optomechanical resonators, and accumulated along several sprays. Sensor performances do not severely degrade after a single deposition event, allowing the same device to be used for multiple detection events.}
	\end{figure*}
\newpage

\section{Mechanical coupling between the disk and the particle}
\begin{figure*}
	\includegraphics[width=0.8\textwidth]{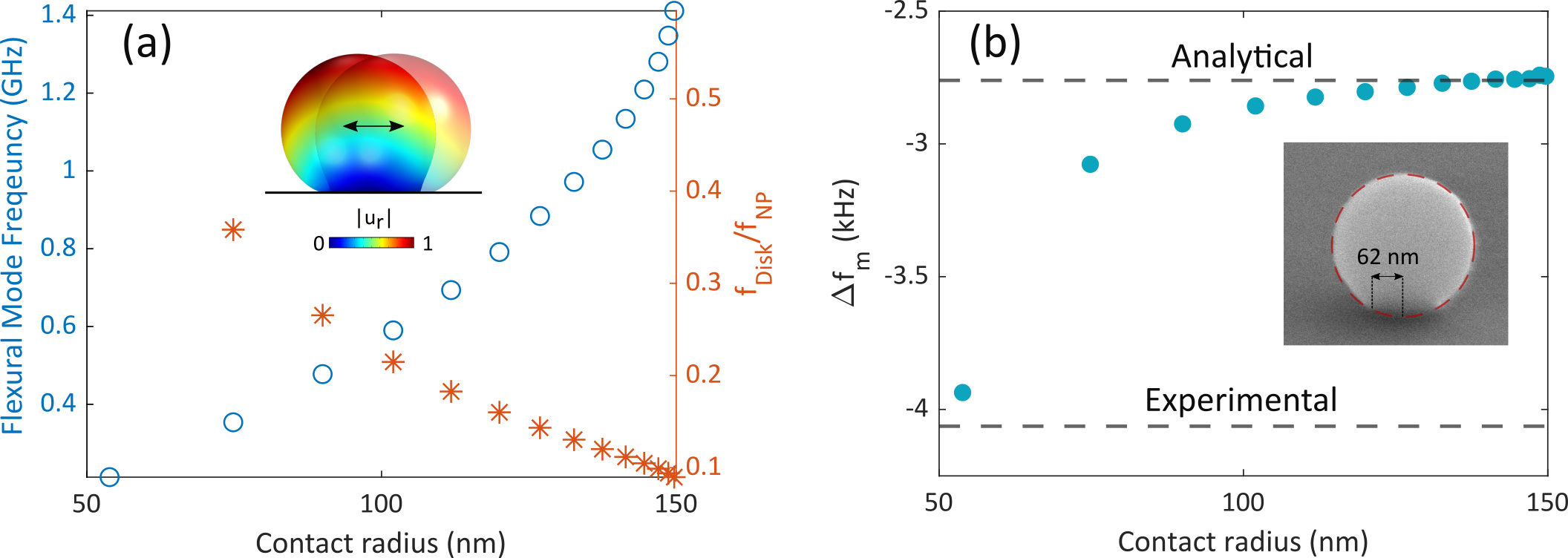}
	\caption{(a) The frequency of the nanoparticle flexural mode, whose mode profile is shown in inset, increases with the contact radius (open blue circles). The ratio between the eigenfrequency of the disk RBM1 and that of this flexural mode is shown in red stars. (b) At small contact radius, when this ratio approaches one, there is a strong deviation between the mechanical frequency shift expected from the point-mass approximation (Analytical) and that found by FEM simulation of the flexural nanoparticle vibration interacting with the disk (filled blued circles). An example of particle contact radius imaged by SEM is reported in inset, together with the associated measured frequency shift (Experimental dashed line).}
	
	\label{Fig:MechanicalCoupling}
\end{figure*}
The vibrations of a nanoparticle attached onto the disk surface were simulated by 3D FEM. The vibration profile of the flexural mode arising in such case is shown in the inset of Fig.~\ref{Fig:MechanicalCoupling}~(a). On the left axis of the plot, we report the flexural mode frequency for a latex nanoparticle of 300~nm diameter, for an increasing contact radius with the underlying surface. The frequency ratio between the disk fundamental radial breathing mode (RBM1) and the nanoparticle flexural mode is reported on the right axis. Since this value parametrizes the mechanical coupling between the disk and nanoparticle vibrational modes, we expect to fall in the "inertial regime" for a ratio smaller than 1 (large contact radius) and to enter the "coupling regime" in the opposite case (ratio close to 1, and small contact radius). In these FEM simulations, the particle/disk contact area has been obtained by removing a spherical cap from the bottom part of the spherical nanoparticle, and compensating the material density of the rest of the nanoparticle in order to preserve its overall mass.

The point-mass approximation (Sauerbrey’s equation 1 of the main text) applies in the inertial regime, whereas it becomes inadequate to estimate the frequency shift induced by mass absorption in the coupling regime. In the latter, mechanical mode splitting and hybridized quality factors between the disk and nanoparticle are instead expected. For instance, FEM simulations in this regime have been performed to calculate the frequency shift induced by the landing of the single latex particle introduced in Fig. 2 of the main text. The results are reported in Fig.~\ref{Fig:MechanicalCoupling}~(b) for a varying contact radius (filled blue circles), and compared to the shift expected from the analytical formula (point mass approximation, Analytical dashed line) for the same mass, and to the shift obtained experimentally (Experimental dashed line). The analytical formula is able to reproduce the FEM results for large contact radii but deviates in the opposite situation. For this specific measured nanoparticle, the contact radius is equal to 62 nm (see inset of Fig.~\ref{Fig:MechanicalCoupling}~(b)), which seems to explain why we measured a frequency shift 46\% larger than the one expected from the point-mass approximation. This deviation is a consequence of the nanoparticle elasticity in our measurement. It can be used to estimate the nanoparticle Young modulus, provided the contact radius and mass of the nanoparticle are known.
\end{document}